\documentstyle[preprint,aps,epsf]{revtex}
\input amssym.def
\input amssym
\tighten
\begin{document}
\draft
%\twocolumn[\hsize\textwidth\columnwidth\hsize\csname @twocolumnfalse\endcsname

\title{A New Mechanism for Leptogenesis} 
\author{R. Jeannerot \\ 
       {\normalsize{ Department of Applied Mathematics and
Theoretical Physics, Cambridge University,}}\\ 
        {\normalsize{ Silver Street, Cambridge, CB3 9EW, UK}}}
\date{\today}
\maketitle
\begin{abstract}
Unified theories containing a $U(1)_{B-L}$ gauge symmetry
predict heavy Majorana right-handed neutrinos. In such theories, cosmic strings 
may form at the $B-L$ breaking scale. If the Higgs field forming the
strings is also the Higgs field which gives mass to the
right-handed neutrinos, there are right-handed neutrinos trapped as 
transverse zero modes in the core of the strings. When cosmic string loops
decay, they release these neutrinos. This is an
out-of-equilibrium process. The released neutrinos acquire heavy
Majorana mass and decay into massless leptons and   
electroweak Higgs particles to produce a lepton asymmetry, which is converted into a baryon asymmetry via sphaleron
transitions. 
\end{abstract}
%\vskip2pc]

%\newpage
 
Grand unified theories (GUT's) provide the
standard scenario for baryogenesis, via the out-of-equilibrium decays
of heavy gauge and Higgs 
bosons which violate baryon number ($B$) and lepton number ($L$). But it
has been realized 
a decade ago that, unless the universe started with a non-vanishing
$B-L$ asymmetry, any $B$ or $L$ asymmetry generated at the GUT scale
would be erased by sphaleron transitions \cite{KRS}. An initial $B-L$
asymmetry can be obtained in theories containing an 
extra gauge ${\rm U}(1)_{B-L}$ symmetry via the out-of-equilibrium  
decays of heavy Majorana
right-handed neutrinos \cite{Y86,Luty,Y93}. This mechanism however
requires either 
very heavy neutrinos or extreme fine tuning of the parameters in the
neutrinos mass matrix \cite{Y93}. Also, the masses of the new gauge
bosons must be bigger than the smallest heavy neutrino mass
\cite{Enqvist}. Hence there is a wide range of parameters for which 
the mechanism does not produce enough baryon asymmetry.

In this letter, we show that in unified models involving an extra
gauge ${\rm U}(1)_{B-L}$ symmetry, a  primordial $B-L$ asymmetry can be
generated by the out-of-equilibrium decays 
of heavy Majorana right-handed neutrinos released by collapsing cosmic
string loops.  
As a consequence of ${\rm U}(1)_{B-L}$ breaking, cosmic strings may
form at the $B-L$ breaking scale, according to the Kibble mechanism
\cite{Kibble}.  
We call them $B-L$ cosmic strings. The
Higgs field mediating the breaking of $B-L$ is the Higgs field forming
the strings and it is the same Higgs field  
that gives heavy Majorana mass to the right-handed neutrinos. Hence,
due to the winding of the Higgs field around the 
string, we expect right-handed neutrino zero modes \cite{Jackiw}
trapped in the core of the strings. 
These zero modes are 
predicted by an index theorem \cite{Weinberg}. There are also
modes of higher energy bounded to the strings. We shall
consider only the zero modes, which are the most
favourable to be trapped. $B-L$ cosmic string loops lose their energy
by emitting 
gravitational radiation and rapidly shrink to a point, releasing
these right-handed neutrinos. This is an out-of-equilibrium
process. Right-handed neutrinos 
acquire heavy Majorana mass and decay into massless leptons and 
electroweak Higgs particles to produce a lepton asymmetry. This lepton
asymmetry is converted into a baryon asymmetry via sphaleron
transitions.

Topological $B-L$ cosmic strings form when a gauge group 
$G \supset {\rm U}(1)_{B-L}$
breaks down to a subgroup $H \not\supset {\rm U}(1)_{B-L}$ of $G$, if
the vacuum manifold $G\over H$ is 
simply connected, that is if the first homotopy group $\pi_1({G\over H})$
is non-trivial. If $\pi_1({G\over H}) =I$ but string solutions still
exist, then embedded strings \cite{Tanmay} form when G breaks down to
H. Embedded strings are stable for a wide range of parameters. In
left-right models, embedded $B-L$ strings usually form. In the simple
U(1) extension of the standard model ${\rm SU}(3)_c \times  
{\rm SU}(2)_L \times {\rm U}(1)_Y \times {\rm U}(1)_{Y'}$ where $Y'$
is a linear combination of $Y$ and $B-L$ topological strings form \cite{paper3}. In
grand  
unified theories with rank greater than five, such as SO(10) or E(6),
$B-L$ cosmic strings may form, 
depending on the symmetry breaking pattern and on the set of Higgs
fields used to do the breaking down to the standard model gauge
group \cite{tom,paper3}. There is a wide
range  
of theories which contain both ${\rm U}(1)_{B-L}$ and $B-L$ cosmic strings.

Consider a unified model with an extra ${\rm U}(1)_{B-L}$ gauge
symmetry and stable $B-L$ cosmic strings. The gauge and Higgs
fields forming the strings will be the $B-L$ associated gauge boson $A'$ and
the Higgs field $\phi_{B-L}$ used to break ${\rm
U}(1)_{B-L}$. Right-handed neutrinos acquire heavy Majorana mass 
via Yukawa couplings to $\phi_{B-L}$. The ${\rm U}(1)_{B-L}$ part
of the theory is described by the Lagrangian  
\begin{eqnarray}
L &=&  {1\over 4} f_{\mu \nu} f^{\mu \nu} + (D_\mu \phi_{B-L})^\dagger
(D^\mu \phi_{B-L}) - V(\phi_{B-L})  \\ 
&& + i \overline{\nu}_L \gamma^\mu D_\mu \nu_L + i \overline{\nu}_R
\gamma^\mu D_\mu \nu_R \\ 
&+& i \lambda \phi_{B-L} \overline{\nu}_R \nu_L^c - i \lambda
\phi_{B-L}^* \overline{\nu}_L^c \nu_R  + L_f \: . \label{eq:Laglept}
\end{eqnarray}
The covariant derivative $D_\mu =
\partial_\mu -i e a^{Y'}_\mu$ where $e$ is the gauge coupling constant
and $Y'$ is a linear combination of $B-L$ and $Y$. 
$\lambda$ is a Yukawa coupling constant, and $V(\phi_{B-L})$ is the
Higgs potential. The spinor $N = \nu_R + 
\nu_L^c$ is a Majorana spinor satisfying the Majorana condition $N^c
\equiv C \gamma_0^T N^* = N$, where $C$ is the charge conjugation
matrix. Hence $N$ has only two independent components, two degrees of
freedom. $L_f$ 
is the fermionic $B-L$ part of the Lagrangian which does not contain
neutrino fields. 

For a straight infinite
cosmic string lying along the $z$-axis, the Higgs field $\phi_{B-L}$ and
the $Y'$ gauge field $a$ in 
polar coordinates $(r,\theta )$ have the form 
\begin{eqnarray}
&&\phi_{B-L} = f(r) e^{i n \theta} \\
&&a_\theta = - n \tau {g(r) \over e r } \hspace{1cm}  a_z = a_r =0 ,
\end{eqnarray} 
where $n$ is the winding number; it must be an integer. Most strings have
winding number $n=1$; strings with winding number $|n| >1$ are
unstable. $\tau$ is the string's generator; it is the normalised ${\rm
U}(1)_{Y'}$ generator. It has different eigenvalues 
for different fermion fields.  
The functions $f(r)$ and $g(r)$ must satisfy the following boundary
conditions
\begin{eqnarray}
f(0) = 0 \hspace{.75cm} & {\rm and} & \hspace{1cm}  f \rightarrow
\eta_{B-L} \hspace{.5cm} {\rm as } \hspace{.5cm} r \rightarrow \infty ,
\\ 
g(0) = 0 \hspace{.75cm} & {\rm and}& \hspace{1cm} g \rightarrow 1
\hspace{.5cm} {\rm as }\hspace{.5cm}  r \rightarrow \infty 
\: ,
\end{eqnarray}
where $\eta_{B-L}$ is the scale of $B-L$ breaking.  The exact
forms of the functions $f(r)$ and $g(r)$ depend on the Higgs potential
$V(\phi_{B-L})$.

From the Lagrangian (\ref{eq:Laglept}) we derive the equation 
for the right-handed neutrino field: 
\begin{equation}
i \gamma^\mu D_\mu \nu_L^c - i \lambda \phi_{B-L}^* \overline{\nu}_L^c
= 0 \: \label{eq:mot1}  
\end{equation} 
where $\nu_L^c = C \gamma_0^T \nu_R^*$ .
Solving (\ref{eq:mot1}), we find that Majorana neutrinos trapped as
tranverse zero modes in the core of $B-L$ cosmic
strings have only one independent component. For an $n=1$ vortex it takes
the form :  
\begin{equation}
N_1 = \beta(r,\theta) \: \alpha(z+t) \label{eq:nuR}
\end{equation}
where $\beta(r,\theta)$ is a function peaked at $r=0$ which
exponentially vanishes outside the core of the string,
so that the fermions effectively live on the strings. The $z$ and $t$
dependence of $\alpha$ shows that the neutrinos travel at
the speed of light in the $-z$ direction, so that they are effectively
massless. In an $n = -1$ vortex, the function $ \alpha = \alpha(t-z)$,
so that the fermions travel at the 
speed of light in the $+z$ direction. These fermions can be described by
an effective theory in $1+1$ dimensions. The usual energy to 
momentum relation 
\begin{equation}
E = P \label{eq:E}
\end{equation}
holds. We have no boundary conditions in the 1 spatial dimension, and
the spectrum of states is continuous. In the ground state the Fermi 
momentum of the zero modes is $p_F = 0$.

The field solution (\ref{eq:nuR}) and the energy to momentum 
relation (\ref{eq:E}) have been
derived for fermions on a straight infinite string. However physical cosmic
strings are very wiggly and are not 
straight. Hence relations (\ref{eq:nuR}) and (\ref{eq:E}) do not hold in 
the physical case. Neither do they hold for 
cosmic string loops, even if the latter are assumed to be smooth. 
On a cosmic string loop, fermions are characterised by  
their angular momentum $L$. The energy 
relation becomes \cite{BarrMathe}  
\begin{equation}
E = {(L+{1\over 2})\over R} = P + {1\over 2R} \label{eq:E2}
\end{equation}
and hence the energy spectrum is 
\begin{equation}
E ={ ({\rm n}  + {1\over 2})  \over R} \: \label{eq:E1} 
\end{equation}
where $n \in {\Bbb N}$. We see from Eqs. (\ref{eq:E}), (\ref{eq:E2}),
and (\ref{eq:E1}) that, when 
$R$ is very large, the string looks locally like a straight string. We
have an almost continuous spectrum of states. The fact that the string gets 
a finite curvature acts as a perturbation on the string bound states. The 
energy levels get quantised and the Fermi energy gets a non-vanishing 
value, $E_F = {1\over 2 R}$. As the string loop shrinks, its radius $R$ 
decreases and we see from Eq.(\ref{eq:E1}) that the Fermi energy increases 
and that the separation between energy levels gets wider.

Assuming that a loop decays when its radius
$R$ becomes comparable to its width $w \sim 
\eta_{B-L}^{-1}$,  we deduce that the Fermi energy level
when the loop decays is $E_F \sim {1\over 2} \eta_{B-L}$, where the
$B-L$ breaking scale $\eta_{B-L}$ is of the order of the  
right-handed neutrino mass. $E_F$ is lower than the energy needed by
the neutrinos to escape the string \cite{BarrMathe}.  Hence, when a
cosmic loop decays, it  
releases at least $n_\nu = 1$ heavy Majorana neutrinos. Quantum
fluctuations and finite temperature corrections may increase
$n_\nu$. Part of the final  
burst of energy released 
by the decaying cosmic string loop is converted into mass energy for
the gauge and Higgs particles released by the string, and into mass energy for
the neutrinos. A decaying $B-L$ cosmic string loop releases heavy
$B-L$ Higgs particles which can decay into right-handed neutrino
pairs, and hence increase $n_\nu$. This is an out-of-equilibrium
process. Due to angular momentum conservation, the massive Majorana
neutrinos released by a decaying cosmic string loop 
which were trapped as transverse zero modes are spinning particles.

Heavy Majorana right-handed neutrinos interact with the standard model
leptons via the Yukawa couplings 
\begin{equation}
L_Y = h_{ij} \overline{l_i} H_{ew} \nu_{Rj} + h.c.
\end{equation} 
where $l$ is the usual lepton doublet; for the first family $l = (e,
\nu)_L$. $H_{ew}$ is the standard model doublet of Higgs
fields. Majorana right-handed neutrinos can decay via the diagrams shown in
Fig. 1.a. and 1.b. CP is violated through the one loop radiative
correction involving a Higgs 
particle as shown in figure 1.b. The right-handed neutrinos are
out-of-equilibrium, and hence a lepton asymmetry can be generated. The
lepton asymmetry 
is characterised by the CP violation parameter $\epsilon$ which,
assuming that the 
neutrino Dirac masses fall into a hierarchical pattern qualitatively
similar to that of the leptons and quarks, is estimated to
be \cite{Luty}  
\begin{equation}
\epsilon \simeq {m_{D_3}^2 \over \pi v^2} {M_{N1} \over M_{N2}} \sin{\delta
  } \label{eq:CP} 
\end{equation}
where $m_{D_3}$ is the Dirac mass of the third lepton generation, $v$
is the vacuum expectation value of the electroweak Higgs field $v = 
<H_{ew}> = 174$ GeV, $M_{N1}$ and $M_{N2}$ are the right-handed
neutrino Majorana masses of the first and second generation respectively and
$\delta $ is the CP violating phase.

The corresponding $B-L$ asymmetry (we use $B-L$ instead of $L$
since the ($B+L$)-violating electroweak anomaly conserves
$B-L$) must be calculated solving
Boltzmann equations which take into account all $B$, $L$ and $B+L$
violating interactions and their inverse decay rates. We can however
calculate the $B-L$ asymmetry produced taking into account only the
out-of-equilibrium decays of 
right-handed neutrinos released by decaying cosmic string loops and
assuming that 
the rates of inverse decays are 
negligible. Hence an upper limit on the baryon number  
 per commoving volume at temperature $T$ is then given by \cite{KolbTurner}
\begin{equation}
B(T) =  {1\over 2} {N_{\nu}(t) \epsilon \over s} , \label{eq:B-L}
\end{equation}
where s is the entropy at time $t$ and 
$N_\nu(t)$ is the number density of
right-handed neutrinos which have been released by 
decaying cosmic string loops at time $t$. Recall that the temperature
$T$ is related to the cosmic time $t$ via the relation  $t = 0.3 \:
g_*^{-{1\over 2}} {M_{pl} \over T^2}$, where $g_*$ counts the number
of massless degrees of freedom in the 
corresponding phase and $M_{pl}$ is the Planck mass. $s$, the entropy
at time $t$, is given by $s = {2\over 45} \pi^2 g_* T^3$. $N_\nu(t)$
is approximately $n_\nu$ times the number density of cosmic string  
loops which have shrinked to a point at temperature $T$. Assuming that
sphaleron transitions are not in thermal equilibrium below $T_{ew}$,
and neglecting any baryon number violating processes which might have
occured below $T_{ew}$, the baryon number of the universe at
temperature $T \leq T_{ew}$ is then given by 
\begin{equation}
B = B (T_{ew}), \label{eq:today}
\end{equation}
which is also the baryon number of the universe today.
If sphaleron transitions are also rapid below $T_{ew}$, we should
include the neutrinos released below $T_{ew}$. However, below $T_{ew}$
the number density of decaying cosmic string loops is negligible, and
hence this 
would not affect the result in any sense.

The number density of decaying cosmic string loops can be estimated from
the three scales  model of reference\cite{ACK}. The model is based 
on the assumption that
the cosmic string network evolution is characterised by three length
scales $\xi (t)$, $\overline{\xi }(t)$ and $\chi (t)$ related to the
long string density, the persistence length along the long strings
(which is related to the fact that the typical loop size is much
smaller than $\xi$), and the small scale structure along the strings
respectively. 

Cosmic string loops lose their energy by emitting
gravitational radiation at a rate  $\dot{E} = - \Gamma_{loops} G
\mu^2$ \cite{ACK}, where $\mu \sim T_c^2$ is the string
mass-per-unit-length and $T_c= 
\eta_{B-L}$ is the critical temperature of the phase transition
leading to the string network formation. $G$ is Newton's 
constant. The numerical factor $\Gamma_{loops}\sim 50 - 100$ depends
on the loop's 
shape and trajectory, but is independent of its length. The mean size
of a loop born at $t_b$ is assumed to be $(k-1) \Gamma_{loops} G \mu
t_b$. At a later time $t$, it is then $\Gamma_{loops} G \mu (k t_b -
t)$.  The loop finally disappears at a time $t = k t_b$. Numerically,
k is found to lie between 2 and 10. 

The rate at which the string loops form in a volume $V$ is given by \cite{ACK}
\begin{equation}
\dot{N} ( t_b) = {\nu V \over (k-1) \Gamma_{loops} G \mu t_b^4}
\end{equation}
where the parameter $\nu$ can be expressed in terms of the various
length scales of 
the model which vary with time. We start with
$\xi  \sim \overline{\xi } \sim \chi $. Then $\xi $ and later
$\overline{\xi}$ will start to grow and will evolve to the scaling
regime characterised by $\xi (t)$ and $\overline{\xi }(t) \sim t
$. The length scale $\chi $ grows much less rapidly. Therefore $\nu $
varies with time. In the scaling regime, $\nu$ is estimated to lie in
the range $\nu = 0.1 - 10$ \cite{ACK}. 

Note that it has recently been shown that
cosmic string networks reach the scaling solution at a time $t_*$ much smaller 
than previously estimated \cite{Martins}. The authors of
ref.\cite{Martins} find that in the radiation dominated, for minimal
GUT strings, $t_{*} = 8 \times 10^2 \, t_c$ \cite{Martins}, where $t_c$ is
the time at 
which the strings form, and the associated temperature is $T_{*}
\simeq 10^{14.5}$ GeV. Hence our approximation for the 
rate of cosmic string loop formation is suitable; it leads to a lower
bound on the number of 
cosmic string loops. Only numerical simulations could lead us to  
a better estimate, but it is beyond the scope of this paper.

Since it has been shown that most of the loops formed have relatively
small size, we shall assume that the number of loops
rejoining the network is negligible and thus that the number of
decaying loops is equal to the number of forming loops. Hence the
number density of right-handed neutrinos which have been released by
decaying cosmic string loops at time $t$ is given by  
\begin{equation}
N_\nu (t) = n_\nu \int_{k t_c}^t {\dot{N}(t_b) \over V} \Bigl({r(t_b)
\over r(t)}\Bigr)^3 \, dt_b 
\end{equation}
where where $r(t)$ is the cosmic scale factor and $n_\nu$ is the mean
number of right-handed neutrinos released 
by a single decaying loop. In the radiation dominated era $r(t) \sim
t^{1\over 2}$. After integration we obtain 
\begin{equation}
N_\nu (t) = {2\over 3} n_\nu \: {\nu \over (k-1) \Gamma_{loops} G
\mu} \: { 1 \over (0.3)^3 g_*^{-{3\over 2}}} \: \Bigl[{1\over k^{3\over 2}}
({T_c 
  \over M_{pl}})^3 - ({T \over M_{pl}})^3 \Bigr] T^3 .
\end{equation}
 Hence the baryon number per commoving volume today produced by the
decays of heavy Majorana right-handed neutrinos released by decaying
cosmic string loops given in Eq.(\ref{eq:B-L}) becomes   
\begin{eqnarray}
B \simeq {7.5  g_*^{1\over 2}\over ( 0.3 \: \pi)^3} \: {\nu \over (k-1)
k^{3\over 2 } \Gamma_{loops}} {T_c 
  \over M_{pl}} \: {m^2_{D_3} \over v^2}\: {M_{N1} \over M_{N2}} \:
\sin{\delta } ,
\end{eqnarray}
where we have used Eq.(\ref{eq:CP}) for the CP violation parameter
$\epsilon$. The produced $B$ asymmetry depends on the cosmic string 
scenario parameters, on the neutrino mass matrix and on the strength
of CP violation. 

We now calculate the lower and
upper bounds on $B$ which correspond to different values of
the parameters in the cosmic string scenario. We fix the
neutrino mass matrix parameters and assume maximum CP violation,
i.e. $\sin { \delta } = 1$. We assume that the Dirac mass of
the third generation fermions lies in the range $m_{D_3} =1 - 100$
GeV, and that the ratio 
of the right-handed neutrino masses of the first and second generation
${M_{N1} \over M_{N2}} = 0.1$. With the above assumptions, we find $B$ to
lie in the range  
\begin{equation}
B \simeq (1 \times 10^{-10} - 5 \times 10^{-2}) \: g_*^{1\over 2} \:
({T_c \over M_{pl}}) , 
\end{equation}
and we see that $B$ strongly depends on the cosmic string scenario
parameters.  Recall that the baryon number-to-photon ratio
${n_B \over n_\gamma}$ is related to the baryon number of the universe
$B$ by ${n_B \over n_\gamma} = 1.80 \, g_* \, B$. Hence, if $g_* 
\simeq 10^2$, our mechanism alone can explain the baryon-number-to-photon
ratio predicted by nucleosynthesis, ${n_B \over n_\gamma} = (2-7)
\times 10^{-10}$,  
with the $B-L$ breaking scale in the range 
\begin{equation}
\eta_{B-L} =  (2 \times 10^{7} - 3 \times 10^{16})  \; {\rm GeV} . 
\end{equation}
We point out that the result could be better
calculated solving Boltzmann equations, which take into account all $B$,
$L$, and $B+L$ violating interactions and do not neglect the inverse decay
rates. Furthermore, the rate of decaying cosmic string loops can be
calculated via numerical simulations, which would have to take into
account the different regimes of the network evolution which occur
during and after the friction dominated era. This may change the
allowed value for the $B-L$ breaking scale by a few orders of 
magnitude. Note also that if CP is not maximally violated, the $B-L$
breaking scale will be shifted towards higher values. Finally, we
recall that when a cosmic string loop decays, it also releases massive
Higgs bosons $\phi_{B-L}$ and massive gauge bosons $a$ which can decay
into right-handed neutrinos. This process is not taken into account
here because the masses of the Higgs and gauge bosons and of the
neutrinos are very close to each other and the Higgs and gauge fields
can also decay into other particles.

\section*{Acknowledgments}

The author would like to thank Brandon Carter, Nick Manton and Mark Hindmarsh 
for discussions. She would like to thank Newnham College and PPARC for
financial support.

\begin{figure} [h]
\epsfbox{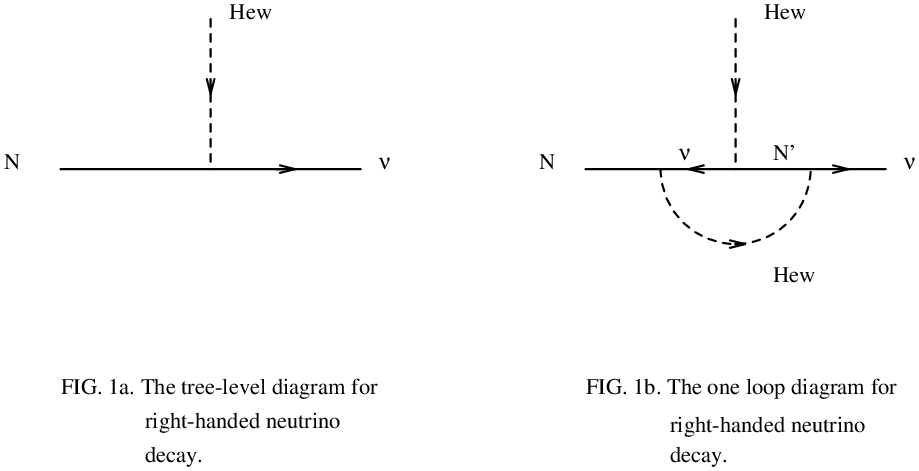}
\end{figure}

\end{document}